 \documentclass[%
 reprint,
 amsmath,amssymb,
 aps,
 5pt
]{revtex4-2}
\usepackage{physics}
\usepackage{mathtools}
\usepackage{amssymb,amsfonts,amsmath}
\usepackage{float} 
\usepackage{booktabs}
\usepackage{xcolor}

\usepackage{graphicx}
\usepackage[colorlinks=true, allcolors=blue]{hyperref}
\usepackage{cleveref}

\usepackage{dcolumn}
\usepackage{bm}

\newcommand*{\rom}[1]{\expandafter\@slowromancap\romannumeral #1@}

\begin{document}

\title{Shear-driven arrest of dip-coating film entrainment}

\author {Theo Witkamp$^1$, Stéphane Zaleski$^2$, and Federico Toschi$^1$}
 
\affiliation{$^{1}$Department of Applied Physics and Science Education,
Eindhoven University of Technology, 5600 MB Eindhoven, Netherlands}
\affiliation{$^{2}$ Sorbonne Universit\'e, CNRS and IUF, UMR 7190, Institut Jean Le Rond d'Alembert, Paris, 75005, France}
 
\date{\today}
\begin{abstract}
A plate withdrawn from a bath of partially wetting liquid entrains a film above a critical velocity. Here, we show that this critical threshold can be substantially increased by the presence of a laminar gas flow opposed to the plate velocity that exerts a tangential shear stress on the liquid–gas interface. Using two-phase immiscible Navier–Stokes simulations and confining the gas to a narrow channel, we derive a closed-form expression for the interfacial stress, which combined with lubrication theory relates the thickness of the film and the contact line velocity to the applied shear stress. As the gas velocity is increased, the film thins and the contact line velocity slows down, until the imposed shear overcomes the drag of the moving plate and entrainment is arrested. As in the quiescent case, the velocity mismatch between plate and contact line is independent of the plate velocity and is set by the applied shear alone. We find that one set of varying gas flow simulations, with constant critical plate velocity, therefore determines the critical gas flow for every smaller plate velocity.
\end{abstract}

\maketitle

\begin{figure}[h!]
    \centering
    \includegraphics[width=1\linewidth]{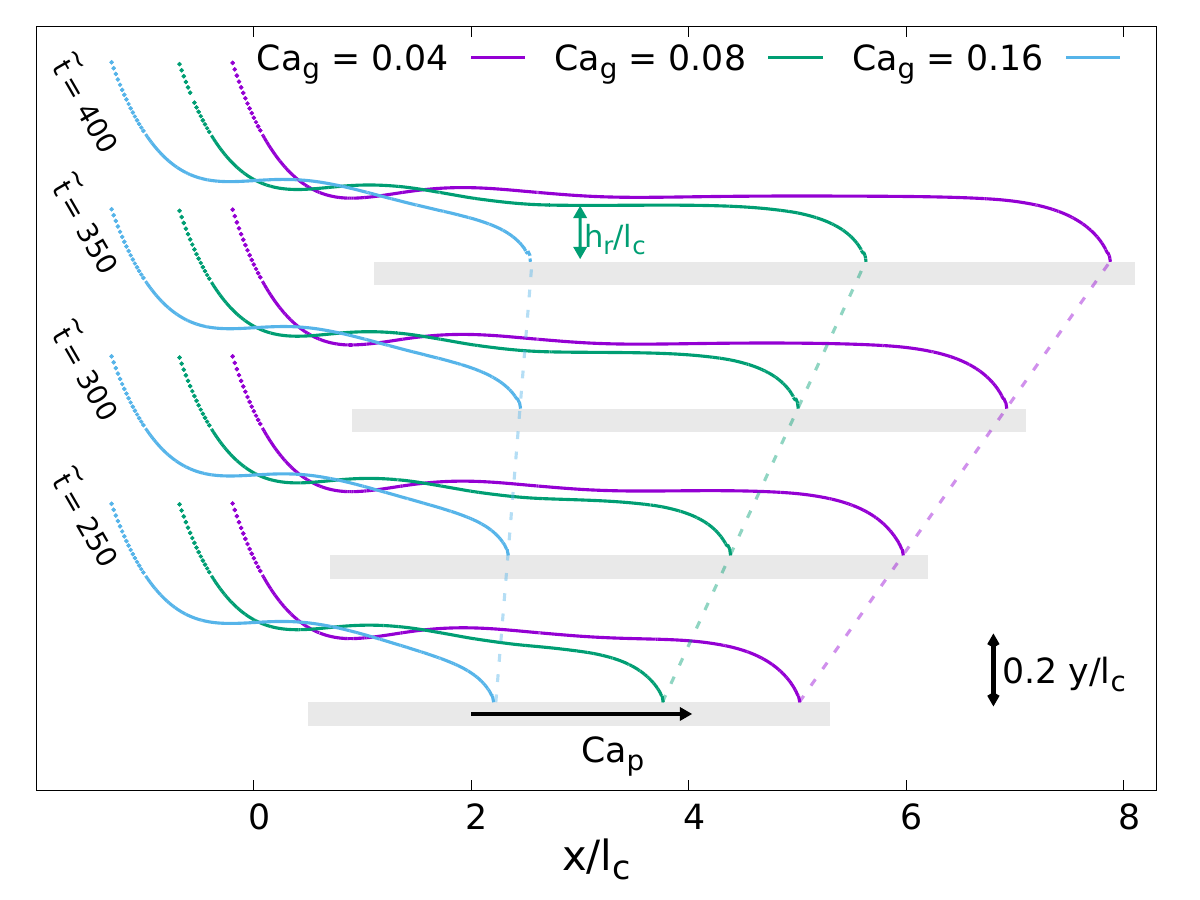}
    \caption{Evolution of the liquid-gas interface, in the steady state, for various gas velocities and times $\tilde t = t \mu l_c /\gamma$, for plate velocity Ca$_p = 0.05$. The dashed line shows the contact line position in time. In the steady state, the velocity is constant, as in the case with no gas flow. The blue interfaces, Ca$_g = 0.16$, corresponding to a near-critical gas flow, show that for strong enough gas capillary number, the contact line motion is almost zero. As Ca$_g$ is increased, the height of the ridge $h_r$ is decreased.}
    \label{fig:RidgeWind}
\end{figure}

    \emph{Introduction} - Wetting films occur in a wide range of natural and industrial processes such as droplets sliding down leaves, galvanization processes where plates are coated by being withdrawn from baths or immersion lithography  where the behavior of the triple contact point between gas, liquid, and solid plays a crucial role~\cite{mulkens2004benefits}. The natural setup to study the motion of thin films on a plate is the Landau-Levich-Derjaguin (LLD) problem, in which a plate is drawn out of a bath and the liquid coats the plate under specific conditions. To understand this wetting transition, the natural dimensionless group, constructed from the viscosity $\mu_l$, the surface tension $\gamma$ and the plate velocity $U_p$ is the capillary number Ca = $\mu_l U_p/\gamma$, which compares viscous stresses $\mu_l U_p/l_c$ against the capillary pressure $\gamma/l_c$ inside the film. Here the capillary length $l_c = \sqrt{\gamma/\rho_l g}$ is the typical length scale associated with this problem. Above the critical capillary number Ca$_c$, the capillary pressure can no longer sustain the viscous drag and the liquid coats the plate~\cite{eggers2004hydrodynamic}. In practice, ~\citet{snoeijer2006avoided} show that this forced-wetting transition is governed by the velocity difference between the plate and the contact line, when the liquid partially wets the surface, i.e. when the contact angle between the solid and the interface is non-zero. In steady state the contact line recedes relative to the plate at a velocity $U_p-U_{cl}$ that is fixed by the local dynamics near the contact line and is independent of the plate velocity, with $U_{cl}$ the contact line velocity. The mismatch feeds a capillary ridge whose flat-film height follows from a lubrication balance, $h_r=l_c\sqrt{3\mu_l(U_p-U_{cl})/\gamma} \propto l_c\text{Ca}^{1/2}$, which matches to the LLD film of height $h_{LLD} \propto l_c\text{Ca}^{2/3}$ through a capillary shock~\cite{snoeijer2006avoided}. A natural extension to the dip-coating setup is to introduce a shear at the liquid-gas interface. For example, the presence of thermal gradients, inducing Marangoni flows, can pull films up a plate, while preserving the characteristic ridge structure near the contact line~\cite{bertozzi1998contact, cazabat1990fingering}. In industrial coating processes, gas jets are widely used to control film thickness by impinging on the entrained liquid film~\cite{aniszewski2020planar,yahyaee2017numerical}. Gas flows also influence droplet detachment under gravity~\cite{white2021wind}, drive rivulet motion, and, at sufficiently high shear rates, trigger hydrodynamic instabilities that lead to film breakup and three-dimensional flow structures~\cite{zhang2021experimental}. Generally, these settings consider gas flows with significant inertia which rely on models to quantify the effects of shear. Here, we consider a gas flow with low inertia, constrained in a channel, which allows us to analytically compute the interfacial shear. Understanding how to delay the wetting transition is relevant to several industrial applications such as immersion lithography \cite{winkels2011receding}.
\begin{figure}[h!]
    \centering
    \includegraphics[width=0.75\linewidth]{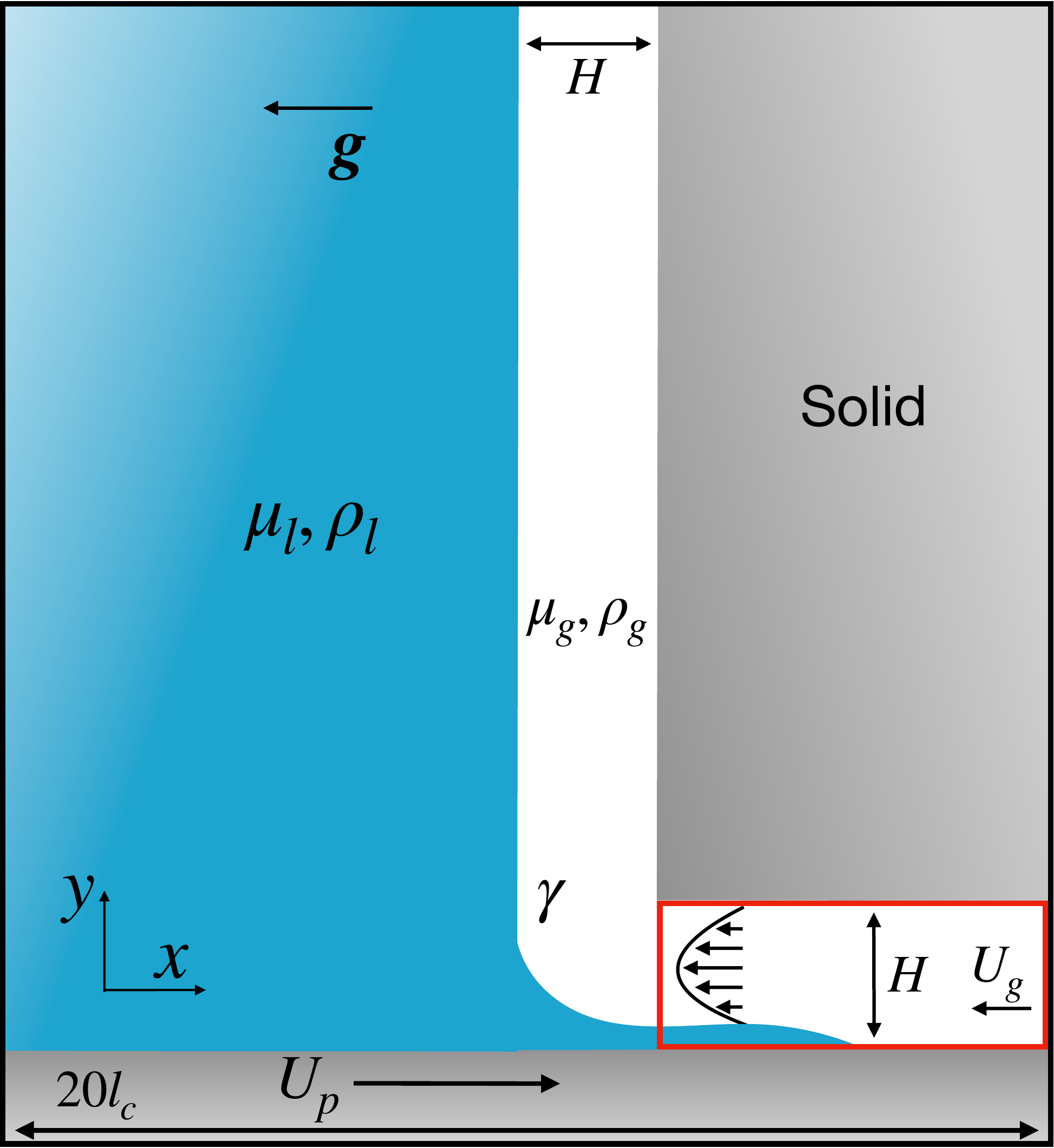}
    \caption{Sketch of the setup. The liquid and gas have viscosities $\mu_l, \mu_g$ and densities $\rho_l, \rho_g$, respectively. The gravity $g$ acts opposite to the motion of the liquid film. The liquid-gas surface tension is $\gamma$. The gas flow is confined in a rectangular channel of height $H$ and enters on the right with a plug velocity $U_g$. The red area is the zone of interest, shown in more detail in Fig. \ref{fig:domzoom}.}  
    \label{fig:domain}
\end{figure}
To study the effect of a shearing gas flow on the contact line dynamics, we use the setup presented in Fig. \ref{fig:domain}. We simulate the liquid and gas phases using a two-phase immiscible and incompressible description of the Navier-Stokes equations. The continuum hypothesis of the Navier-Stokes equations is particularly convenient for analytical and numerical purposes. One consequence, however, is the divergence of the stress near the contact line where the liquid, gas, and solid intersect, under the no-slip boundary condition~\cite{huh_hydrodynamic_1971}. Physically, molecular effects take place at distances $l_s \sim \mathcal{O}(10^{-9}\text{m})$ from the wall. Additionally, several physical effects, such as evaporation, van der Waals forces or surface roughness, influence the dynamics of the contact line\cite{lacis2020steady, snoeijer_moving_2013}. In our framework, we only consider the length scale $l_s$ which can be introduced in the continuum description with a Navier slip condition\cite{lauga2003effective}. Under these assumptions, we find that applying a shear at the interface of the entrained film through the use of a gas flow reduces both the contact line velocity and the height of the ridge. For sufficiently high shears $\tau$, the motion of the contact line can be stopped for a given plate velocity $U_p$. We also show that the mismatch between the plate and contact line velocity $U_p - U_{cl}$ remains independent of the plate velocity and is only a function of the applied shear. A direct consequence is that a single set of simulations with varying gas flows is sufficient to infer the effect of the shear for lower plate velocities. 
\newline
\emph{Numerical setup} - The Navier-Stokes equations are solved using the free and open-source software \emph{Basilisk}~\cite{popinet_accurate_2009, popinet2018numerical,popinet2015quadtree}, where a one-fluid formulation combined with the volume of fluid method is used to simulate the two phases~\cite{tryggvason2011direct}. Dip-coating has already been simulated using Basilisk to understand the effect of grid size on the critical plate velocity~\cite{afkhami_transition_2018} due to the implicit nature of the slip length or to study galvanization processes~\cite{aniszewski2020planar}. As shown by ~\citet{snoeijer2006avoided}, the dynamics of entrainment are well described in two dimensions and edge effects are negligible in the center. We use stress-free boundary conditions for the velocity and a constant hydrostatic pressure on the top and right sides of the domain. The velocity of the plate $U_p$ is imposed tangentially on the bottom wall, with an additional no-penetration condition. The simulations are performed in a domain of size $20 l_c$, discretized with $2^{10}$ grid cells in each direction, leading to the capillary length being discretized by approximately 50 grid cells. We use adaptive mesh refinement with maximum refinement along the interface to ensure that the slip length remains constant during the entire simulation. ~\citet{afkhami_transition_2018} have shown that a domain bigger than $\approx 10 l_c$ is sufficient to properly capture the dynamics. We use a viscosity ratio $\chi = \mu_g/\mu_l$ of $2\cdot10^{-2}$ and a density ratio $\Gamma = \rho_g/\rho_l$ of $10^{-3}$, to model air-water systems. Four additional dimensionless numbers are required; the plate capillary number Ca$_p$ = $\mu_l U_p/\gamma$, the Laplace number La = $\gamma\rho_l H/\mu_l^2$, the Bond number Bo = $\rho_l g H^2 / \gamma$ and the gas capillary number Ca$_g$ = $\mu_g U_g /\gamma$. For all the simulations, the Bond number is Bo = 4.84 and the Laplace number is La = 2.2. The gas flow is confined in a channel of height $H = \sqrt{Bo}l_c \gg h_r$. The channel walls are modeled using an embedded boundary~\cite{ghigo2021conservative} with no-slip and no-penetration boundary conditions. In order to model the gas, a uniform inlet velocity $U_g$ is imposed which develops into a Poiseuille flow, with volumetric flux $Q_{II} = U_gH$ (see Fig. \ref{fig:domain}). The steady Poiseuille profile is achieved far enough from the contact line, such that entrance effects are negligible. The velocity of the gas flow is varied and the Reynolds number inside the channel is Re$_g$ = Ca$_g$La $\chi^2 / \Gamma$ $\sim 1$. 
\begin{figure}[h!]
    \centering
    \includegraphics[width=1\linewidth]{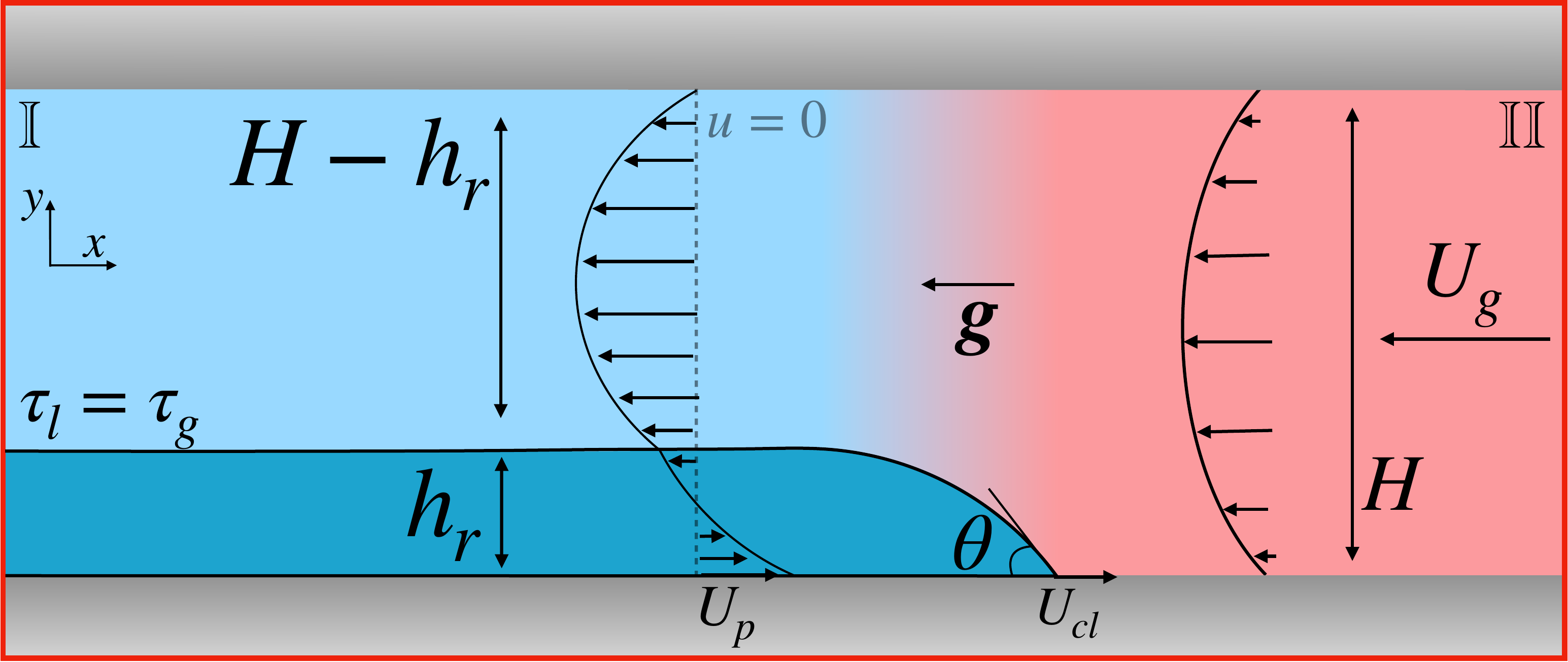}
    \caption{Enlarged view of the channel in Fig. \ref{fig:domain}. A plug flow is imposed on the right of the channel with flux $Q_{II} = U_g H$. In the steady state, the film has height $h_r$ and is flat. Area I is located above the flat-film while Area II is the dry area. Conservation of flux enforces $Q_I = Q_{II}$. At the liquid-gas interface, in the flat-film region, continuity of shear is present, $\tau_l = \tau_g$.}  
    \label{fig:domzoom}
\end{figure}
We impose a static contact angle $\theta_s = 50^\circ$, which is chosen to match ~\citet{snoeijer2006avoided}, where they also use lubrication theory. The contact angle sets the orientation of the interface in the first grid cell, which is then used in the computation of the curvature~\cite{afkhami2008height}. We do not use a dynamic contact angle model since the investigated quantities result from mass-conservation relations and are not affected by assumptions made at the contact line~\cite{snoeijer2006avoided}. The stress divergence near to the contact line is regularized with an implicit slip length $l_s$~\cite{afkhami_transition_2018}, which only acts locally. The implicit nature of the slip is due to the contact line interface being advected by fluxes computed half a gridsize, $\Delta/2$, away from the solid boundary, which results in $l_s \approx \Delta/2$. At distances greater than several grid cells from the contact line, the no-slip boundary condition is recovered. Although the assumptions made at the contact line do not affect the results discussed later, the contact line velocity $U_{cl}$ depends on the slip length and therefore on the refinement. 
\newline 
\emph{Results} - The evolution of the interface is seen in Fig. \ref{fig:RidgeWind} for several inlet velocities $U_g$. The interface profiles are shown in the steady state where the velocity of the contact line $U_{cl}$ is constant, as in the case without gas flow~\cite{snoeijer2006avoided}. Near the contact line, a ridge of height $h_r$ forms and connects to the bath through a dimple. Increasing the gas velocity $U_g$ has the double effect of reducing both the contact line velocity $U_{cl}$ and the height of the ridge $h_r$. 
To study the dynamics of the film as a function of inlet velocity, we use the dimensionless lubrication theory, derived from the incompressibility and long-wave limit Stokes equations~\cite{oron_long-scale_1997, bonn_wetting_2009}
\begin{equation}
    \partial_t h + \partial_x Q_f = 0, 
    \label{eq:consh}
\end{equation}
\begin{equation}
    Q_f = \frac{1}{3}\gamma h^3\partial_x \kappa - \frac{1}{3} \rho_l g h^3 + \frac{\tau}{2} h^2 + \mu_l U_p h, 
    \label{eq:lub_h}
\end{equation}
where $Q_f = \int_0^{h} u \,dy$ is the flux in the film and, $\kappa$ and $\tau$ are the curvature and shear stress evaluated at the liquid-gas interface, respectively. From left to right, the terms on the RHS represent capillarity, gravity, shear at the interface, and drag by the moving plate, respectively. 







\begin{figure}
    \centering
    \includegraphics[width=1\linewidth]{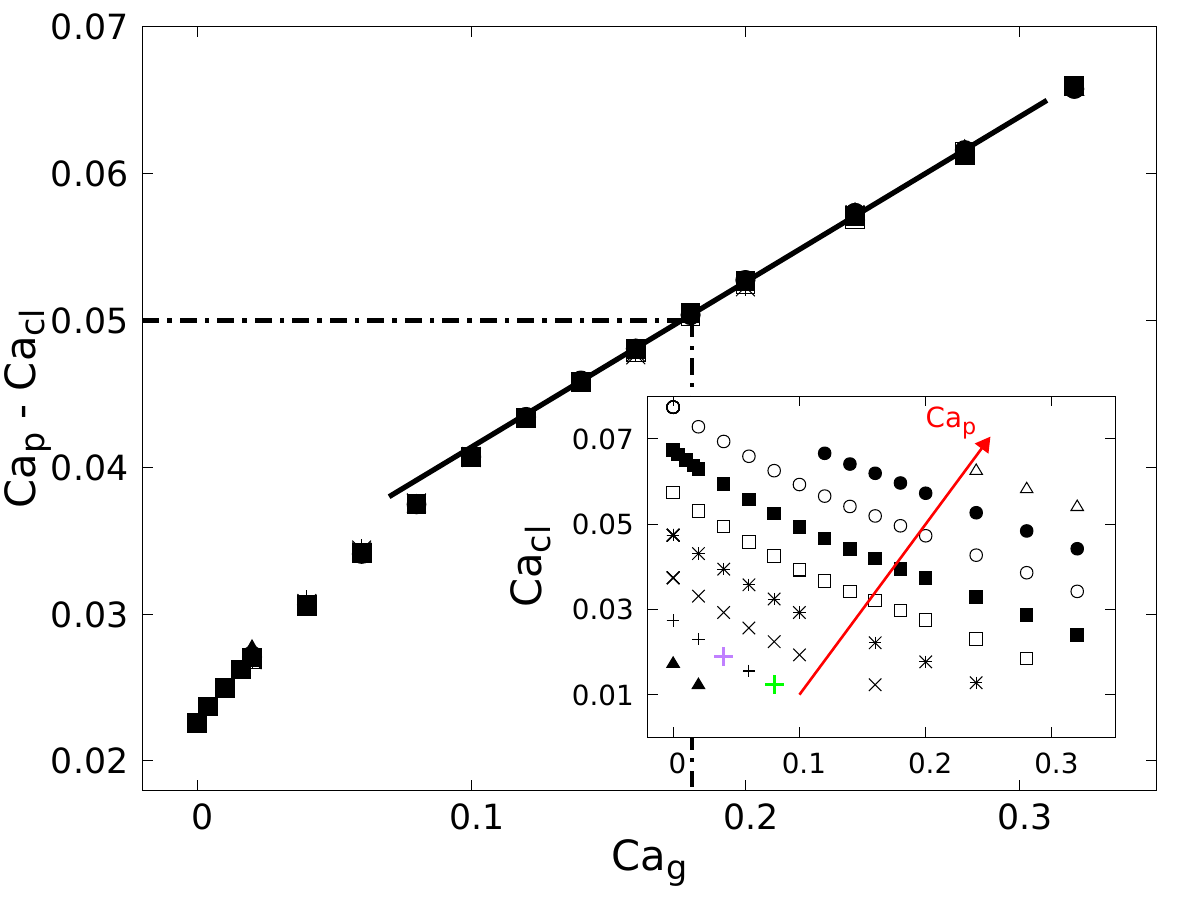}
    \caption{Main panel: The difference between the plate capillary number Ca$_p$ and contact line capillary number Ca$_{cl}$ depends only on the gas capillary number Ca$_g$. The simulations are performed with a maximum resolution of $l_c \approx 50\Delta$. The solid black line shows the linear increase in contact line velocity offset for higher gas velocities. The dash-dotted line shows the critical Ca$_g^* \approx 0.18$ for Ca$_p = 0.05$. In principle, any plate velocity greater than Ca$_p = 0.05$ can be used to compute this value. Inset: The contact line velocity Ca$_{cl}$ is reduced as a function of gas flow. The symbols correspond to constant plate velocity Ca$_p$. The purple and green crosses are the contact line velocities of the corresponding colors in Fig. \ref{fig:RidgeWind}.}
    \label{fig:UpUclvsCaw}
\end{figure}

\begin{figure}[h!]
    \centering
    \includegraphics[width=1\linewidth]{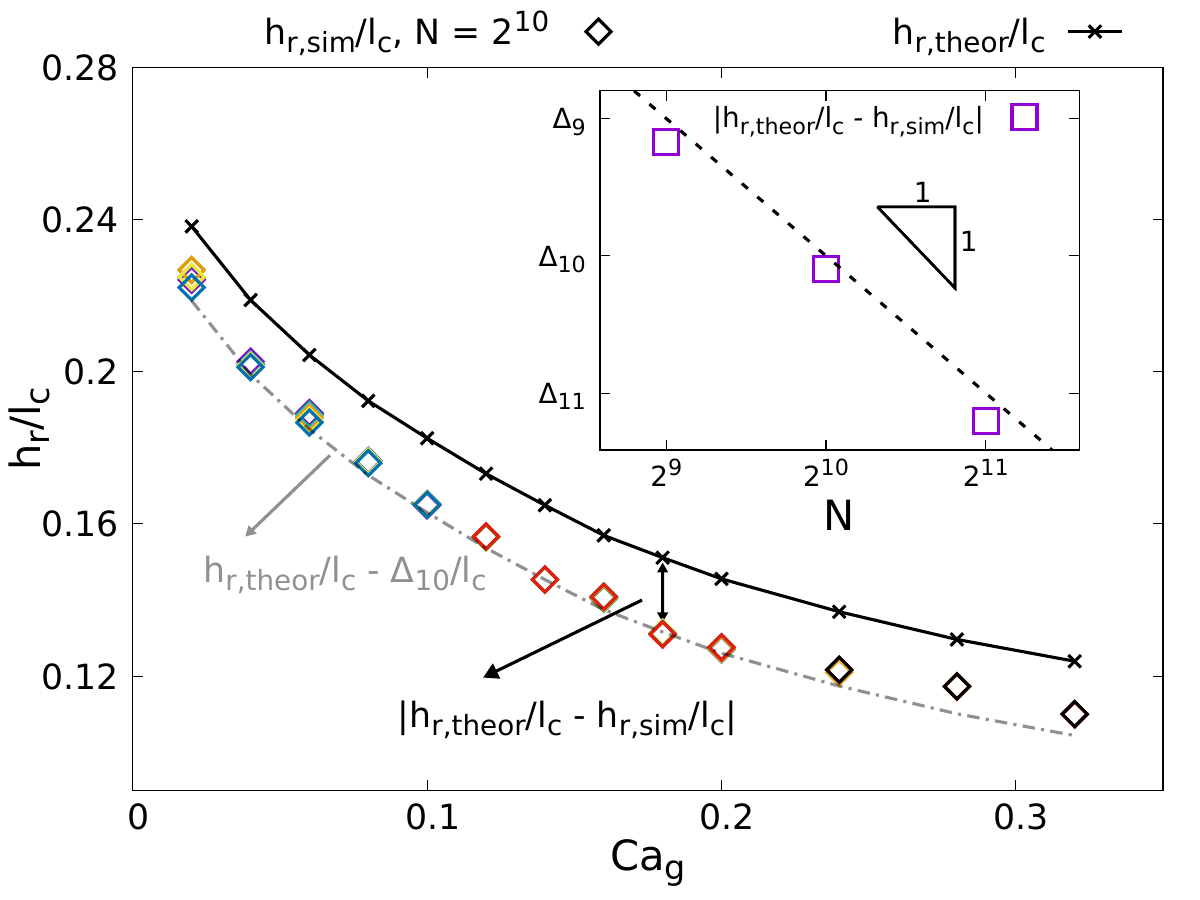}
        \caption{Film height $h_r$ as a function of gas capillary number Ca$_g$. The theoretical prediction is computed using the computed contact line velocities (see Fig. \ref{fig:UpUclvsCaw}). The diamonds indicate the height of the film measured at $x/l_c = 6$, once $x_{cl}/l_c = 8$ and different colors represent different plate velocities. The difference between the predicted height and the grid size ($h_{r,theor} - \Delta_{10}$) is plotted in gray and aligns with the measured heights, showing that the offset is constant as a function of Ca$_g$. The inset shows the convergence of the difference between the theoretical and measured height. In this inset, the vertical coordinate $\Delta_M$ is the grid size corresponding to $N = 2^M$. The difference is explained by the smoothing of the interface necessary to get converging simulations (see text).}
    \label{fig:hcaw_lvl10}
\end{figure}

\begin{figure}[h!]
    \centering
    \includegraphics[width=0.95\linewidth]{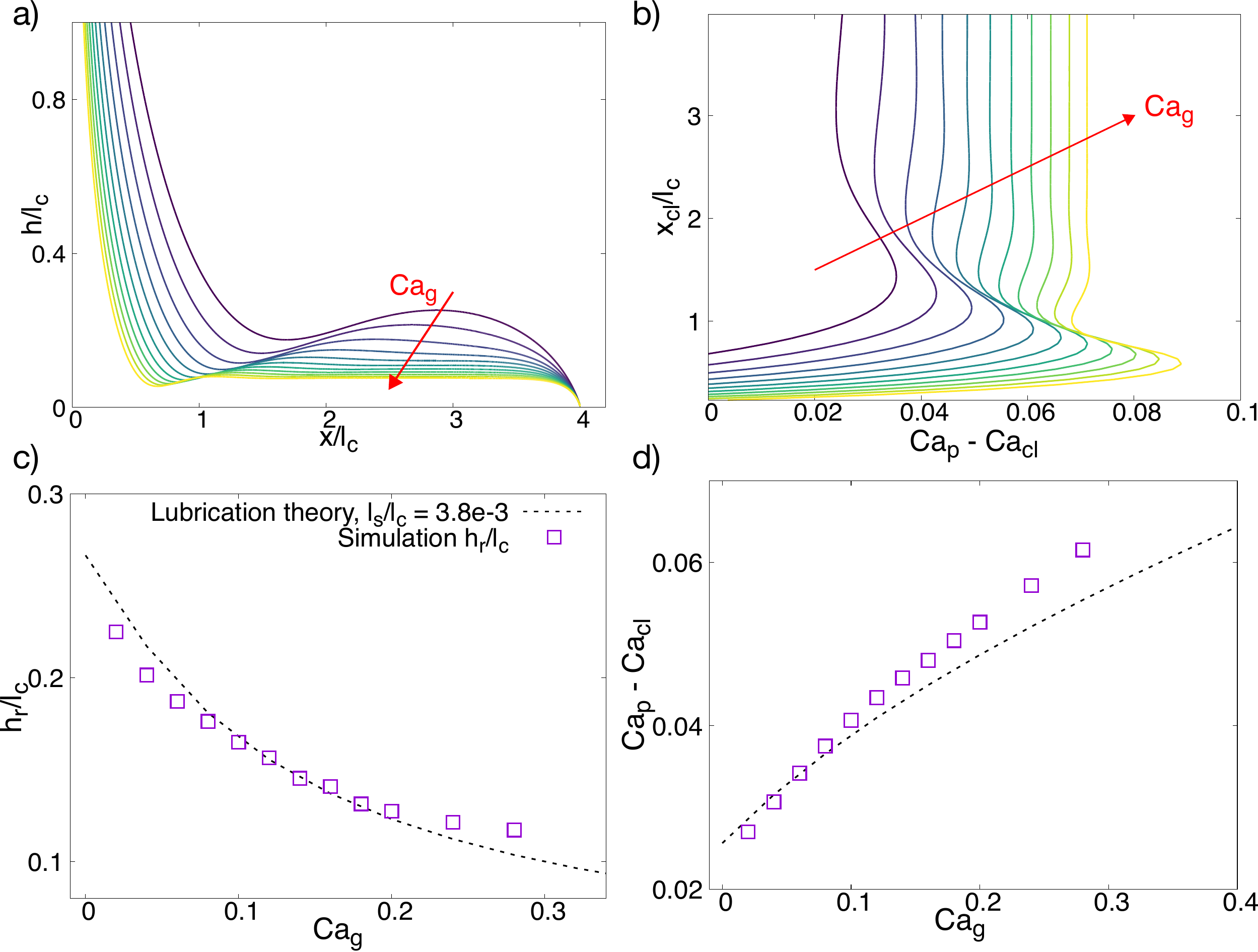}
    \caption{a) Interface profile for several gas flows $U_g \in [0:20]$ for $x_{cl}$ = 4.  b) Bifurcation diagrams for several inlet velocity $U_g$. As the velocity is increased, the transition to wetting occurs at higher plate velocities. c) Height of the ridge $h_r$ measured at $x = 6$ for solution to Eq. \ref{eq:lub_h_em}, where $x_{cl}$ = 8, as a function of $U_g$. d) Transition velocity $U_p - U_{cl}$ as a function of $U_g$.}
    \label{fig:lubtheory}
\end{figure}

\citet{snoeijer2006avoided} show that the flux through the ridge must be $Q_f = U_{cl} h_r$. In the flat-film region, the capillary pressure is zero and the height $h_r$ is in steady state, i.e., $\partial_t h_r = 0$. These assumptions allow us to simplify Eqs. (\ref{eq:consh}) and (\ref{eq:lub_h}) to an algebraic relation 
\begin{equation}
    -\frac 1 3 \rho_l g h_r^2 + \frac{\tau_l}{2}h_r + \mu_l(U_p - U_{cl}) = 0,
    \label{eq:h}
\end{equation}
where $\tau_l$ is now the shear stress evaluated at the ridge height, $\tau_l = \tau(y = h_r)$. To evaluate $\tau_l$, we make use of the continuity of tangential stress at the liquid-gas interface, $\tau_l = \tau_g$, where $\tau_g$ is the interfacial shear exerted from the gas. As seen in Fig. \ref{fig:domzoom}, the gas is confined in an effective channel of height $(H-h_r)$. The boundary conditions to compute the gas velocity profile are: conservation of flux in region I and II, $\int_{h_r}^H u\,dy = U_gH$ and zero-velocity boundary conditions on the top of the channel and at the liquid-gas interface. The latter is motivated by $U_g \sim \mathcal{O}(10)$ being much larger than the interfacial velocity, which is of the order of $U_p$, i.e. $\mathcal{O}(10^{-2})$. From this, the interfacial shear is given by
\begin{equation}
    \tau_l = \tau_g = -\frac{6\mu_g U_g H}{(H-h_r)^2}.
    \label{eq:tau}
\end{equation}
Importantly, Eqs. (\ref{eq:h}) and (\ref{eq:tau}) provide two equations for two unknowns, e.g. $h_r$ and $\tau_l$. The third unknown, $U_{cl}$, depends upon and can be fixed by prescribing an inflow condition from the bath to the film. This relation, which involves several flow and geometrical parameters, can be read out from the numerical simulations. These constraints imply that the mobility law $U_{cl} = f(\theta, H, h_r, ...)$ cannot  be determined from asymptotic matching, as in the classical setup. For qualitative matching of the results, we show in the end matter that solving Eq. \ref{eq:lub_h} with a shear directly proportional to the inlet velocity $\tau \propto U_g$, provides good agreement with the measured contact line velocity and height of the ridge. To verify the validity of Eq. \ref{eq:h}, we will predict the height of the film $h_r$ using the computed contact line velocity $U_{cl}$. 
\newline
\emph{The contact line velocity} - The dependence of the contact line velocity $U_{cl}$ on the inlet velocity $U_g$ is shown in Fig. \ref{fig:UpUclvsCaw}. The inset shows that $U_{cl}$ decreases as $U_g$ is increased, with the curves being separated by a constant offset for different plate velocities. This is confirmed in the main panel, where the curves collapse when considering the mismatch between the plate and contact line velocities ($U_p - U_{cl}$). The mismatch is constant as a function of plate velocity and depends only on $U_g$, which is consistent with the results in the quiescent case~\cite{snoeijer2006avoided}. The black line indicates a linear trend $U_{cl} \propto U_g$ for high gas velocities, this is to be expected since the gas Reynolds number Re$_g$ is $\mathcal{O}(1)$.
In this regime, the shear at the interface is directly proportional to the velocity of the gas. For stronger interfacial shears, the drainage due to gravity becomes negligible. For low inlet velocities, gravity also contributes and a nonlinearity is observed in the contact line velocity dependence. Since the contact line velocity mismatch is independent of the plate velocity, $U_p - U_{cl} = f(U_g)$, measurements performed at a single plate velocity can be used to infer the dynamics at lower plate velocities. Importantly, a critical inlet velocity $U_g^*$ can be defined as 
\begin{equation}
    f(U_g^*) = U_p,
\end{equation}
where $U_g^*$ is the critical gas velocity for plate velocity $U_p$. 
\newline
\emph{Film height} - To avoid channel entrance or exit effects, we let the film progress well inside the channel, until the contact line position reaches $x_{cl}/l_c = 8$. The height is then measured at $x/l_c = 6$, where the curvature of the interface is negligible. The height of the film is independent of the plate velocity, as expected from Eq. (\ref{eq:h}). A monotonic decrease is observed, which is expected from the conservation of mass, where the flux decreases due to both the contact line velocity $U_{cl}$ and the height of the ridge $h_r$. Taking $U_{cl}$ from the simulations, the height of the film predicted by Eq. (\ref{eq:h}) is compared with the simulation results. The agreement is good and the behavior matches the predicted height (see Fig. \ref{fig:hcaw_lvl10}). 
The offset is constant as a function of inlet velocity and is proportional to the grid size. The inset shows the difference between the measured height and theoretical height as a function of refinement. The offset is a numerical artifact of the method used to resolve the interface in the numerical simulations. Because of the high density and viscosity ratios, smoothing the fields across the interface is necessary to get better convergence~\cite{jain2019secondary}. The density and viscosity jumps are averaged within a one-cell neighborhood on both sides of the interface. Since the model in Eqs. (\ref{eq:h}) and (\ref{eq:tau}) assumes a constant viscosity in each phase, we expect the measured height $h_r$ to not align with the theoretical expectation. As the mesh is refined however, the ratio of the filtered region to the film height vanishes and the measured and predicted heights agree. This is confirmed in the inset of Fig. \ref{fig:hcaw_lvl10} where the difference between the measured and predicted height scales directly with the grid size. 
\newline
\emph{Conclusion} - We have shown that a simple gas flow shearing the interface of a film being pulled up a plate greatly affects its height and velocity. As in the quiescent case, the velocity mismatch $U_p - U_{cl}$ remains independent of $U_p$ and depends only on the gas flow velocity $U_g$. A flat-film lubrication model with tangential stress at the interface is used to predict the height and velocity of the contact line. The film height $h$ is reduced by up to a factor of two in the studied regime. We also observe stopping of the contact line motion for sufficiently strong gas flow velocities, for plate velocities which were beyond critical in the quiescent case. Not discussed in this paper is the effect of the gas flow on the meniscus and the connection between the film and the bath. We expect this to be strongly dependent on the geometry of the channel. Future studies could focus on higher Re$_g$, where stagnation pressure and other hydrodynamic instabilities would start to play a role. Other geometries will change the results quantitatively but we expect the qualitative behavior to remain the same. An interesting direction could be the study of the optimal confinement area shape in order to stop the film entrainment.  
\newline
\newline
\emph{Acknowledgments} - We thank Jacob Maarek and Anton Darhuber for the helpful discussions. This work is supported by the Netherlands Organization for Scientific
Research (NWO) through the use of supercomputer facilities (Snellius) under Grant No. 2023.026, 2025.008. This work is funded by the project "High-Velocity Contact Line Dynamics for High-Speed Immersion", which is jointly financed by ASML, Eindhoven University of Technology and by Holland High Tech with the supplementary grant PPS23-2-03536303 for research and development in the topsector HTSM.
\newline
\newline
\bibliography{bib}
\clearpage
\newpage
\onecolumngrid
\begin{center} 
  \textbf{End Matter}
\end{center}
\twocolumngrid
To quantify the effects of a laminar gas flow on the dynamics of an entrained film, we numerically solve 
\begin{equation}
    \gamma\partial_x \kappa - \rho_l g + \frac{3(h + 2l_s)}{2h(h+3l_s)}  \tau+ \frac{3\mu_l(U_p - U_{cl})}{h(h+3l_s)} =0, 
    \label{eq:lub_h_em}
\end{equation}
using an effective shear 
\begin{equation}
    \tau = -\frac{6\mu_gU_gH}{(H-\bar{h}_r)^2}, 
\end{equation}
with $\bar{h}_r/l_c = 0.2$, an estimate for the height of the ridge, independent of the gas velocity, which is needed to get converging simulations. Although the shear only has this form at the interface of the ridge, we expect this to be a reasonable estimate for the magnitude of the stress at the interface and the dependence $\tau \propto U_g$ is correct for laminar gas flows. Note that in the main text, we do not use a slip length in Eq. \ref{eq:lub_h} due to the local nature of the slip in the numerical simulations. Here, the regularization length $l_s$ is needed to avoid the singularity at $h = 0$. To match the height of the ridge $h_r$ in the quiescent case, we find that $l_s/l_c =3.8\cdot10^{-3}$, which is close to the implicit prediction $l_s/l_c \approx \Delta/2 = 10^{-2}$. Fig. \ref{fig:lubtheory}a) shows static solutions of Eq. \ref{eq:lub_h_em} for several gas velocities $U_g$. As the velocity $U_g$ is increased, the ridge height decreases. In panel b), we see that increasing the gas flow increases the mismatch between the plate and contact line velocity, in the steady state. Panel c) and d) compare the predictions of $h_r$ and $U_p - U_{cl}$ resulting from Eq. \ref{eq:lub_h_em} with the measured values from the simulations. We expect the differences to be related to the use of a constant interfacial shear, with the averaged height $\bar{h}_r$.

\end{document}